\documentclass[%
reprint,
 amsmath,amssymb,
 aps,
prl,
]{revtex4-1}

\usepackage{graphicx}
\usepackage{dcolumn}
\usepackage{bm}

\newcommand{\reals}{{\mathbb R}}

\newcommand{\ba}[1]{\begin{array}{#1}}
\newcommand{\ea}{\end{array}}

\newcommand{\be}{\begin{equation}}
\newcommand{\ee}{\end{equation}}
\newcommand{\bea}{\begin{eqnarray}}
\newcommand{\eea}{\end{eqnarray}}
\newcommand{\beann}{\begin{eqnarray*}}
\newcommand{\eeann}{\end{eqnarray*}}

\newcommand{\map}{{f}}
\newcommand{\sden}{\rho}
\newcommand{\cross}{\pi}

\def\reff#1{(\ref{#1})}

\begin{document}

\preprint{APS/123-QED}

\title{Conformal invariance of the 3D self-avoiding walk}

\author{Tom Kennedy}
 \affiliation{Mathematics and Physics Departments, 
University of Arizona.}
 \email{tgk@math.arizona.edu}

\date{\today}

\begin{abstract}
We show that if the three dimensional
self-avoiding walk (SAW) is conformally invariant, then 
one can compute the hitting densities for the SAW in a half space 
and in a sphere. We test these predictions by Monte Carlo simulations and 
find excellent agreement, thus providing evidence that the SAW is 
conformally invariant in three dimensions.
\end{abstract}

\pacs{Valid PACS appear here}
\maketitle


In two dimensions the self-avoiding walk (SAW) is believed to 
be conformally invariant, and this leads to a rich  set
of predictions. In particular the scaling
limit of the SAW in a simply connected domain between 
two fixed points is predicted to 
be the Schramm-Lowener evolution with $\kappa=8/3$
\cite{lsw_saw}. For SAW's in a domain which start 
at a fixed point but are allowed to end anywhere on the boundary, 
there are predictions for the distribution of the terminal 
point \cite{lsw_saw}. 
Simulations of the SAW have found excellent agreement with 
these predictions \cite{Kennedya,Kennedyb,kennedy_lawler}.

In two dimensions the extensive predictions for the SAW can be seen 
as a consequence of the large group of conformal transformations.
In three and higher dimensions this group is rather modest; it is generated by 
Euclidean symmetries and inversions in spheres. Nonetheless it is still
nontrivial, and the use of conformal invariance to study 
critical systems in more than two dimensions has a long history.
In particular, conformal invariance was used to study the SAW in 
\cite{cardy1984conformal,cardyredner}.
So it is natural to ask if the scaling limit of the 
SAW is conformally invariant in three and more dimensions, and if 
such invariance would yield any information about the SAW.  
Since this scaling limit has been proved to be Brownian motion in more 
than four dimensions \cite{hara_slade_a,hara_slade_b}
and substantial progress 
toward proving this for four dimensions has been made \cite{bs},
this question is most interesting in three dimensions. 
In this paper we show that despite the rather limited set of 
conformal transformations in three dimensions, 
conformal invariance allows one to make 
some nontrivial predictions. We test these predictions
by simulations of the SAW and find excellent agreement.

The probability of a SAW $\omega$ is taken to be 
proportional to $\mu^{-N}$ where $N$ is the number of steps 
and $\mu$ is the lattice connectivity constant for the SAW, i.e., 
the reciprocal of the critical fugacity. 
$N$ is not constrained, and there are a 
variety of grand canonical ensembles that we will use. 
We can consider all SAW's in the full space that go between 
two fixed points. We can consider all SAW's in a domain $D$ with 
a variety of possible constraints on the endpoints. 
In the chordal ensemble the two endpoints are fixed points on the 
boundary of $D$. In the radial ensemble one endpoint is a fixed point 
on the boundary and the other is a fixed point in the interior. 
Finally, we can consider an ensemble in which one endpoint is a fixed
point in the interior but the other endpoint can be anywhere on the 
boundary of the domain. 
In this ensemble the location of the endpoint on the boundary is 
random, and we refer to its density as the hitting density.
In all these ensembles the SAW is initially defined on a lattice with 
spacing $\delta$, and we let $\delta \rightarrow 0$ to obtain the 
scaling limit. 
For background on the SAW we refer the reader to 
\cite{des1990polymers}, and for the SAW near a surface to
\cite{eisenriegler1993polymers}.

We will use several critical exponents for the SAW.
Let $N$ denote the number of steps in 
the SAW $\omega$, so $\omega(N)$ is the endpoint of the SAW.
The exponent $\nu$ characterizes the growth of the SAW with the 
number of steps. 
Letting $< \quad >$ denote  expectation with respect to the 
uniform probability measure on SAW's with $N$ steps starting at 
the origin,  $<||\omega(N)||^2> \sim N^{2 \nu}$.
The number of SAW's with $N$ steps starting at the origin
grows as $\mu^N N^{\gamma-1}$ where $\mu$ depends on the lattice 
but the exponent $\gamma$ does not. 
If we only allow SAW's that stay in a half-plane ($d=2$) or 
a half-space ($d=3$), then it grows like $\mu^N N^{\gamma-1-\rho}$.
($\gamma-\rho$ is often denoted by $\gamma_1$.)
So the probability that a SAW with N steps in the full space
lies in the half space goes like $N^{-\rho}$.
The final exponent we will need is the boundary scaling exponent 
$b$. In the chordal ensemble of SAW's between two fixed points 
on the boundary, the partition function will go to zero like 
$\delta^{2b}$ as the lattice spacing $\delta \rightarrow 0$. 
The exponent $b$ is related to the field theory exponent $\eta_\parallel$
that describes the decay of spin-spin correlation along the boundary
by $\eta_\parallel=2b$.
Another characterization of $b$ in two dimensions is that
the probability the SAW will pass through a slit of width 
$\epsilon$ in a curve will go as $\epsilon^{2b}$ as the width goes to zero.
In three dimensions the probability the SAW will pass through a small 
hole in a surface will go as $l^{2b}$ where $l$ is the linear size
of the hole. So it will go as $\epsilon^b$ if $\epsilon$ is the area.
A well known scaling relation between these four exponents 
in $d$ dimensions is $2 b \nu =  2\rho-\gamma  + d \nu $.
A non-rigorous derivation for the SAW may be found in \cite{lsw_saw}.

In two dimensions there are exact, but unproven, 
predictions for these exponents:
$\nu=3/4$  \cite{flory}, $\gamma=43/32$ \cite{nienhuis}, 
$\rho=25/64$ \cite{cardy1984conformal} and $b=\eta_\parallel/2=5/8$ 
\cite{cardy1984conformal}.
In three dimensions there are numerical estimates but no exact predictions: 
$\nu=0.587597(7)$ \cite{clisby_nu},
$\gamma=1.15698(34)$ \cite{gamma_schram}, 
$\gamma_1=0.679 \pm 0.002$ where  
$\gamma_1=\gamma-\rho$ \cite{hegger_grassberger}. 

The chordal and radial ensembles of the SAW are believed to be 
conformally invariant. The ensemble in which the SAW starts at a point 
in the interior and ends anywhere on the boundary is not. 
The hitting density for this ensemble is conformally covariant. 
Since this density is uniform when the SAW starts at the center
of a disc, this conformal covariance completely determines the hitting 
density for simply connected two dimensional domains. 
We give a derivation of the Lawler, Schramm, Werner
formula \cite{lsw_saw} for the conformal covariance of the 
hitting density in two dimensions since our results in three dimensions
will follow the same argument. 

\begin{figure}[tbh]
\includegraphics{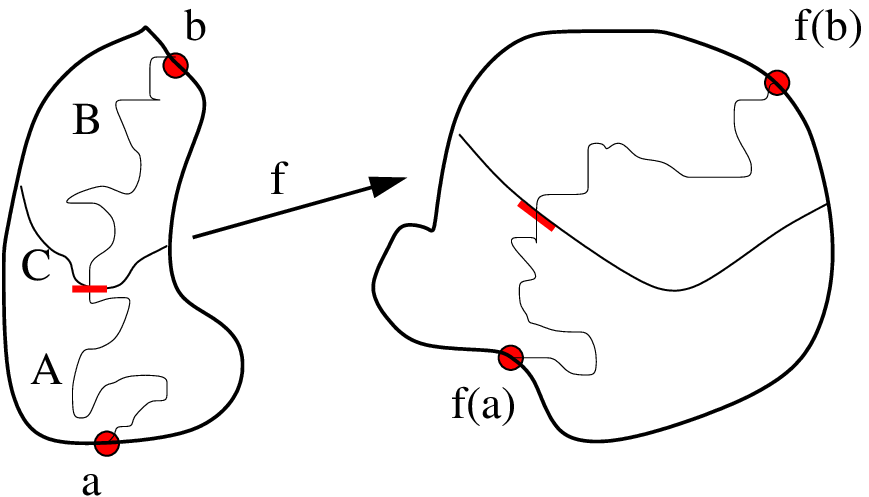}
\caption{\label{fig_hit}}
\end{figure}

Let $D$ be a simply connected domain. Let $C$ be a simple curve between
two boundary points. It divides $D$ into two subdomains which we 
call $A$ and $B$. Let $a$ be a boundary point of the subdomain $A$
which is not on $C$, and $b$ a boundary point of the subdomain $B$, 
not on $C$. See fig. \ref{fig_hit}.
We consider the SAW in $D$ from $a$ to $b$ and condition
on the event that it crosses $C$ only once. This is conditioning on 
an event with probability zero, so we must do it by a limiting 
process. We look at the event that there is a segment of width $\epsilon$ 
in $C$ such that the curve hits $C$ only in this segment. Then 
we let $\epsilon \rightarrow 0$. The probability of passing through a slit of 
width $\epsilon$ should go to zero like $\epsilon^{2b}$.
The prefactor will depend on where we are along the curve and it is 
this prefactor that we interpret as the unnormalized probability that 
the SAW crosses $C$ at that point given that it crosses $C$ only once. 
We denote it by
\beann
\cross_{D,C,a,b}(z)= \lim_{\epsilon \rightarrow 0} 
\epsilon^{-2b} \, P(\epsilon \, slit \, at \, z)
\eeann
where $\epsilon \, slit \, at \, z$ stands for the event that the SAW 
passes through a slit of width $\epsilon$ centered at a point $z \in C$. 

Now let $\map(z)$ be a conformal map on $D$. By conformal invariance, 
the probability the SAW in $D$ goes through the slit in $C$ at $z$ is 
the same as the probability that the SAW in $\map(D)$ from $\map(a)$ 
to $\map(b)$ goes through a slit in $\map(C)$ at $\map(z)$ of width 
$|\map^\prime(z)| \epsilon$. This change of the width of the slit is 
crucial. Equating the two probabilities and canceling the common factor
of $\epsilon^{2b}$, we have 
\bea
\cross_{D,C,a,b}(z) = 
\cross_{\map(D),\map(C),\map(a),\map(b)}(\map(z)) \, |\map^\prime(z)|^{2b}
\label{transform}
\eea

Summing over all SAW's in $D$ that go through a slit should be the same 
as the product of the sum over all SAW's in $A$ from $a$ to the slit times
the sum over all SAW's in $B$ from $b$ to the slit. Let 
$\sigma_{A,C,a}(z)$ denote the hitting density for a SAW in $A$ which 
starts at $a$ and ends at $z$ in the boundary arc $C$. Then we 
should have 
\beann
\cross_{D,C,a,b}(z) \propto \sigma_{A,C,a}(z) \, \sigma_{B,C,b}(z)
\eeann
Combining this with \reff{transform} yields the conformal 
covariance for the hitting density,
\beann
 \sigma_{A,C,a}(z) \propto  \sigma_{\map(A),\map(C),\map(a)}(\map(z)) \, 
|\map^\prime(z)|^b
\eeann
Depending on how one defines the ensemble of SAW's that end on the boundary,
there is a multiplicative correction to this formula that comes from 
lattice effects that persist in the scaling limit \cite{kennedy_lawler}.

This argument works in three dimensions with the caveat that the only 
conformal transformations $f$ are the transformations generated by 
translations, rotations, dilations and inversions in spheres.
We consider the conformal transformation
$
\map(x,y,z) =\frac{2(x,y,1-z) }{ x^2 + y^2 + (1-z)^2}
\label{cmap}
$
It maps $\infty$ to the origin, maps the origin to 
$(0,0,2)$ and maps the unit sphere centered at the origin to the plane 
$z=1$. Note that this is the plane that bisects
the line segment between the two endpoints, so we will 
refer to it as the bisecting plane.  
We parametrize the unit sphere with spherical coordinates $\theta,\phi$. 
Then 
$\map( \cos(\phi) \, \sin(\theta),\sin(\phi) \, \sin(\theta),\cos(\theta))
= (u,v,1)$ with 
$u = \sin(\theta) \cos(\phi)/(1-\cos(\theta))$, 
$v = \sin(\theta) \sin(\phi)/(1-\cos(\theta))$.
The map $\map$ takes 
an infinitesimal area $\epsilon$ on the sphere to an infinitesimal area
$J(\theta,\phi) \, \epsilon$ on the plane $z=1$ with 
$J(\theta,\phi)=1/(1-\cos(\theta))^2$. 
In the $u,v$ coordinates $J(u,v)=(u^2+v^2+1)^2/4$. 

Now consider the scaling limit of the SAW from the origin to $\infty$ in 
$\reals^3$. We consider a small area $\epsilon$ on 
the sphere and condition on the event that the 
SAW intersects the sphere only inside this area. 
This probability goes to zero as $\epsilon^b$ as
$\epsilon \rightarrow 0$, and by symmetry it is independent of where 
the area is on the sphere. 
The map $\map$ takes this to a SAW between $(0,0,0)$ and $(0,0,2)$ 
conditioned to hit the bisecting plane only inside an area of size 
$J(u,v) \epsilon$ on the plane.
Equating these two probabilities, 
$\epsilon^b = \epsilon^b J(u,v)^b \cross(u,v)$ where
$\cross(u,v)$ is the unnormalized probability density for 
where the SAW crosses the plane when we condition on the event 
that it only crosses it once.  
A SAW that crosses this plane only once can be thought of as the 
concatenation of a SAW in the half space $z<1$ from the origin to 
some point on the plane and a SAW in the half space $z>1$ from the point
$(0,0,2)$ to the same point on the plane. So $\cross(u,v)$ should 
be the product of the hitting densities for these two half-space 
SAW's. By symmetry these two hitting densities are equal, and we denote 
them by $\sigma_1(u,v)$. Thus we have  $\sigma_1(u,v) \propto J(u,v)^{-b/2}$.
The constant of proportionality is determined by this
being a probability density.
If we consider the scaling limit of the SAW in the half-space $z<a$ 
starting at the origin and ending on the plane $z=a$, then  
by scaling the hitting density is 
\bea
\sigma_{a}(u,v) \propto [u^2+v^2 + a^2]^{-b} 
\label{hit_plane}
\eea

We now consider a sphere of radius $1$ centered at the origin, and let
$\sden_a(\theta,\phi)$ be the hitting density for 
a SAW inside this sphere which starts at the point $(0,0,a)$ 
and ends on the surface of the sphere.
If we take a SAW in the full space from $(0,0,a)$ to $\infty$ 
and condition on the event that it crosses the sphere only once, 
then the probability density for where it crosses the sphere should 
be just $\sden_a(\theta,\phi)$ since the hitting density for the 
portion of the SAW from the sphere to $\infty$ will be uniform
on the sphere. 
Now we use the same conformal map $f$. 
It takes $(0,0,a)$ to $(0,0,c)$ where $c=2/(1-a)$. 
The probability the SAW starting at $(0,0,a)$ 
and going to $\infty$ goes through a small area $\epsilon$ 
on the unit sphere 
is proportional to $\sden_a(\theta,\phi) \epsilon^b$.
This small area is mapped to an area $J(\theta,\phi) \, \epsilon$
on the plane $z=1$. The probability the SAW from $(0,0,c)$ to 
$(0,0,0)$  goes through this area should be proportional to 
$J(\theta,\phi) \epsilon^b$ times the product of the hitting densities
for two SAW's which end at the same point on the plane with one 
in the half space $z<1$ starting at $(0,0,0)$ and  
the other the half space $z>1$ starting at  $(0,0,c)$.
Thus
$
\sden_a(\theta,\phi) \propto J(\theta,\phi)^b
\sigma_1(u,v) \sigma_{c-1}(u,v)
$.
Using \reff{hit_plane} and expressing $u,v$ in terms of $\theta,\phi$, we 
find after some algebra that
\bea
\sden_a(\theta,\phi) \propto
\left[1 + a^2 - 2a \cos(\theta) \right]^{-b} 
\label{hit_sphere}
\eea
Note that $[1 + a^2 - 2a \cos(\theta)]^{1/2}$ is just the distance from the 
point on the sphere to $(0,0,a)$. If we take $b=3/2$, then 
\reff{hit_plane} and \reff{hit_sphere} are the hitting densities for
Brownian motion in the two geometries. This is not surprising since 
the above argument can also be applied to the ordinary random walk 
for which $b=d/2$. 


There is another prediction about the SAW that follows 
trivially from conformal invariance. 
Consider the point where a SAW in the full space from 
the origin to $\infty$ first hits the unit sphere. This point will
be uniformly distributed over the sphere. 
After the conformal transformation $f$, this becomes the first hit
of the plane $z=1$ for a SAW in the full space from $(0,0,0)$ to 
$(0,0,2)$. 
Its distribution will be the image of the uniform measure 
on the sphere under $\map$. 
So its density with respect to Lebesgue measure on the plane is 
\bea
\frac{1}{\pi}  (u^2+v^2+1)^{-2}.
\label{hit_bisect}
\eea

The most efficient algorithm for simulating the SAW is the pivot 
algorithm whose speed has been dramatically improved recently \cite{clisby}.
It simulates SAW ensembles with 
a fixed number of steps. One endpoint of the SAW
is fixed, but the other endpoint is not. In the 
ensembles we wish to study the number of steps is not fixed, and either 
both endpoints are fixed or one is fixed and the other 
is constrained to lie on some boundary. 
So we cannot directly test the predictions of 
conformal invariance with the pivot 
algorithm. Instead we use the pivot algorithm to study 
ensembles that we expect to have the same scaling limit as the 
ensembles we wish to study, but which are amenable
to simulation by the pivot algorithm. The idea behind these ensembles 
was introduced in \cite{kennedy_dilation}.
Details may be found in \cite{Kennedy_3d_long}.

We first consider the ensemble of SAW's in the full space
between the origin and a fixed point $P$.  
We simulate it using the ensemble of SAW's in the full space 
with $N$ steps which start at the origin and end anywhere. 
Given such a SAW we apply a Euclidean symmetry (rotation and dilation)
that fixes the origin and take the other endpoint of the SAW to $P$. 
If we give these transformed SAW's equal weight, then this will 
not approximate the ensemble we want. However, as we argue in
\cite{Kennedy_3d_long}, if we weight the transformed $N$ step SAW's $\omega$ 
by $||\omega(N)||^{-\gamma/\nu}$, then as $N \rightarrow \infty$ we will obtain 
the scaling limit of the usual ensemble of SAW's between the origin and $P$.

\begin{figure*}
\includegraphics{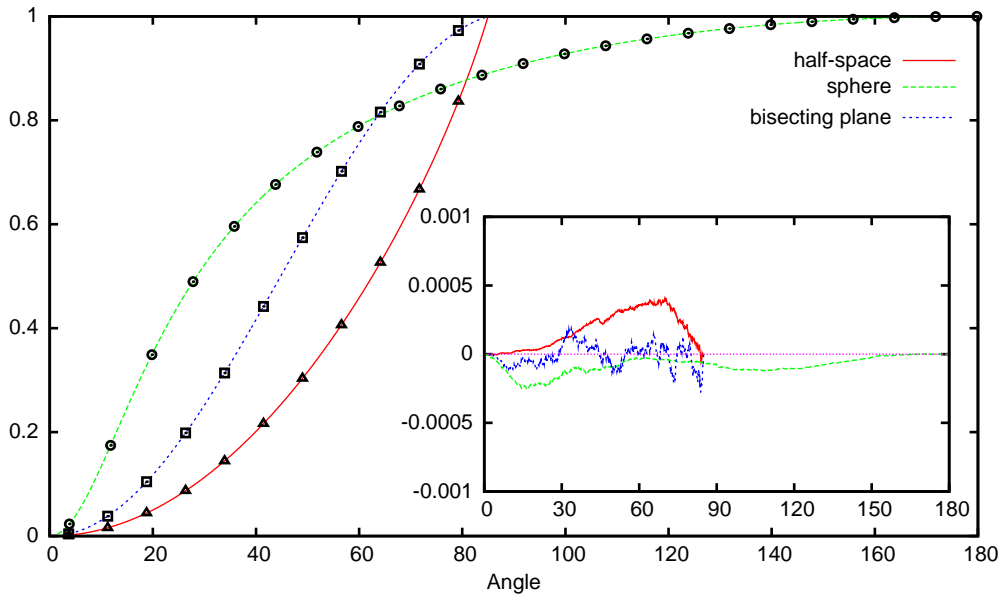}
\caption{\label{fig_all}
The curves are the predicted cdf's for the hitting density for the
half-space (eq. \reff{hit_plane}) and the sphere (eq. \reff{hit_sphere})
and for the first hit of the bisecting plane (eq. \reff{hit_bisect}). The 
circles are the simulation results. The inset shows the differences of 
the predicted and simulated cdf's.
The distance from the center of the sphere to the starting point of the 
SAW is $a=3/4$.
}
\end{figure*}

To simulate the ensemble of SAW's in the half space $z<1$ that start 
at the origin and end anywhere on the plane $z=1$, 
we use the ensemble of $N$-step SAW's starting at the origin 
that stay in the half-space $z>0$. For any such SAW we can dilate it 
and translate it to produce a SAW in the half-space $z<1$ that goes
between the origin and the plane $z=1$. We argue in \cite{Kennedy_3d_long}
that if we weight these SAW's by $||\omega(N)||^{-(\gamma-\rho)/\nu}$, 
then in the scaling limit we get the desired ensemble. 

Finally, to simulate the ensemble of SAW's in the sphere of radius 
$1$ centered at the origin which have one endpoint at 
$(0,0,a)$ and the other endpoint anywhere 
on the surface of the sphere, we 
start with the ensemble of $N$ step SAW's in the full space that start 
at the origin. We dilate the walk to 
to produce a walk with one endpoint at the origin and the other endpoint 
on the sphere of radius $1$ centered at $(0,0,-a)$. 
We then condition on the event that 
this dilated walk lies entirely inside the sphere. As we argue in 
\cite{Kennedy_3d_long}, if we weight these walks by a suitable function of 
$\omega(N)$, then as $N \rightarrow \infty$ we will 
get the scaling limit of the desired ensemble. 
In this ensemble there are lattice effects that come from 
the orientation of the surface of the sphere with respect to the 
lattice. They persist in the scaling limit and 
must be taken into consideration when we test our prediction for 
the hitting density \cite{Kennedy_3d_long}.

We simulate all three of these ensembles 
with $N=10^6$ and generate on the order of $100$ million samples.
The simulation of the SAW in the sphere 
took a little more than $100$ CPU-days. 
The other two simulation were much faster, taking on the 
order of $100$ CPU-hours.
When the SAW $\omega$ ends close to where it starts, 
$||\omega(N)||$ will be relatively small. Such SAW's are improbable, 
but the weighting factor is large, and so the 
statistical errors in the simulation are increased. 
Such walks are excluded from the ensemble of SAW's in the sphere, but
are a problem in the other two ensembles. 
The troublesome SAW's typically 
have a value of $\theta$ near $90$.
So if we condition the random variable 
$\theta$ on $\theta \le \theta_0$ with $\theta_0$ less than $90$,  
we can reduce this problem. 
We take $\theta_0=85$. 

The results of our simulations testing these predictions are shown in 
figure \ref{fig_all}. For the SAW in the sphere we take the distance
from the center to the starting point of the SAW to be $a=3/4$. 
For all three of the predictions we study the 
cumulative distribution function (cdf) of the random variable $\theta$.
In the main figure the three curves are the predictions for the cdf's and 
the points give the simulation results at selected values of $\theta$.  
They are indistinguishable in this figure.  
The differences of the simulation cdf's and the predicted cdf's are 
shown in the inset.
Note the scale of the vertical axis. 
The differences are on the order of a few hundredths of a per cent, 
thus providing strong evidence that the SAW is 
conformally invariant in three dimensions.

{\it Acknowledgements:} 
The support of the Mathematical Sciences Research Institute where this 
research was begun is gratefully acknowledged.

\bibliography{saw_3d_prl_revised}

\end{document}